\documentclass[structabstract]{aa}
\usepackage{graphicx}
\usepackage{txfonts}
\begin{document}
   \title{The Paczy{\'n}ski-Wiita potential}

   \subtitle{A step-by-step ``derivation''}

   \author{Marek A. Abramowicz
          \inst{~}
          }

   \institute{Dept. Physics, G{\"o}teborg Univ.,
   412-96 G{\"o}teborg, Sweden;
             Copernicus Astronomical Ctr., Bartycka 18, 00-716 Warszawa,
             Poland\\
             \email{Marek.Abramowicz@physics.gu.se}
             }

   \date{Received ????; accepted ????}


  \abstract
   {Paczy{\'n}ski realized that a properly chosen gravitational
   potential may accurately model (in a ``pseudo Newtonian'' theory)
   general relativistic effects that determine motion of matter
   near a non-rotating black hole. Paczy{\'n}ski's choice,
   known today as the ``Paczy{\'n}ski-Wiita potential'', proved to be very
   practical. It was used by numerous researchers in the black
   hole accretion theory, and became a standard tool in relativistic
   astrophysics.}
   {The model is an example of Paczy{\'n}ski's admired ability to invent
   ``out of nowhere'' simple ideas that were brilliant, deep and
   useful. Paczy{\'n}ski has guessed intuitively the form
   of the potential. However, it could be also derived by a
   a step-by-step formal procedure. I show the derivation here.}
   {My derivation is based on a standard definition of the
   relativistic ``effective potential'' in the Schwarzschild spacetime.}
   {The relativistic effective potential may be uniquely divided into its
   ``gravitational'' and ``centrifugal'' part. The gravitational part differs
   from the Paczy{\'n}ski-Wiita potential only by a constant.}
   {}


   \maketitle


 \section{A brief history of the potential}

 \cite{polish-doughnuts}, working in Pa\-czy{\'n}\-ski's
 research group in Warsaw, found a practical mathematical
 scheme to construct fully relativistic models of thick
 accretion disks, known today as ``Polish doughnuts''.
 The models displayed several astrophysically interesting features,
 among them seriously super-Eddington luminosities, long and narrow
 funnels that could collimate jets, and a self-crossing ``Roche
 lobe'' equipotential. The scheme developed in Warsaw was suitable
 for both analytic and numerical calculations.

 Some properties of the thick disks were obviously
 connected to strong-field effects of general relativity. Our leader
 Bohdan Paczy{\'n}ski, who was not familiar with the technicalities
 of general relativity, asked me to find a Newtonian way to describe
 these effects. I was rather unhappy about Paczy{\'n}ski's request,
 because initially I wrongly imagined that the only method adequate for
 the task should be the post-Newtonian scheme. It uses tedious, long and
 boring expansions. I was desperately working, producing
 longer and longer formulae, when one day Paczy{\'n}ski came to my
 office, and said ``Stop working on that. I found the solution.''
 And he showed me his solution --- a Newtonian\footnote{It is
 often called ``pseudo Newtonian'' to stress that it does not
 obey the Poisson equation. However, when the external gravity is fixed,
 (as it is in the black hole accretion theory) ``pseudo Newtonian'' is
 practically equivalent to ``Newtonian''. This is why I am using
 both terms here.} potential,
 \begin{equation}
 \label{pacz-wiita-potential}
 \Phi(r) = -\frac{GM}{r -r_{\rm G}},~~ ~~~r_{\rm G} = \frac{2GM}{c^2},
 \end{equation}
 where $r$ is the spherical radius, $M$ is the mass of the black hole,
 and $r_{\rm G}$ is its gravitational radius.
 Paczy{\'n}ski had checked that the two most important
 radii that characterize circular Keplerian orbits, the radius of
 marginally stable orbit $r_{ms}$ (i.e. ISCO) and the radius of
 marginally bound orbit $r_{mb}$ have the same values in
 Newton's gravity with his potential (\ref{pacz-wiita-potential})
 as in Einstein's gravity in the Schwarzschild metric,
 \begin{equation}
 \label{isco}
 r_{ms} = 3\,r_{\rm G},~~~r_{mb} = 2\,r_{\rm G}.
 \end{equation}
 This was a brilliant display of the qualities of Paczy{\'n}ski's mind:
 he just {\it guessed} the right, simple and powerful solution
 to the problem. His solution immediately proved to be very
 practical. Shortly afterwards, \cite{pacz-wiita} used
 (\ref{pacz-wiita-potential}) to numerically calculate
  models of thick disks. The models differed from these calculated
 with the full strength of general relativity by only
 a few percent. Later, this opened a flood gate: numerous
 authors used the Paczy{\'n}ski-Wiita potential in their
 calculations of black hole accretion flows.
 The potential is so remarkably successful, that some researchers
 use it even outside its obvious limits of applicability:
 (a) for rotating black holes; and this is wrong because
 (\ref{pacz-wiita-potential}) does not include the
 Lense-Thirring effect, (b) for self-gravitating fluids; and
 this is wrong because $\nabla^2\Phi \not = 0$.


 \section{A ``derivation'' of the Paczy{\'n}ski-Wiita potential}

 Why is the Newtonian Paczy{\'n}ski-Wiita potential
 (\ref{pacz-wiita-potential}) such an accurate model of the strong
 relativistic effects? Should this be considered a fortunate,
 unexpected coincidence, or could one ``derive'' the
 potential from the first principles of Einstein's general
 relativity? I remember discussing this question shortly
 with Thibault Damour in the late 1970. Although we were
 convinced that the ``effective potential approach'' should
 provide such a derivation, we have not completed relevant
 calculations. I summarize them here.

 In Newtonian theory, let $E$ be energy, $L$ angular momentum,
 $\Phi(r)$ gravitational potential, and $V$ radial velocity.
 The orbital motion is often described in terms of the effective
 potential $U(r, L) = \Phi(r) + L^2/2r^2$,
 \begin{equation}
 \label{newton-effective-equation}
 \frac{1}{2}\,V^2 = E - U(r, L),
 \end{equation}
 Circular orbits locate at the effective potential extrema,
 \begin{equation}
 \label{circular-Newton}
 \left( \frac{\partial U}{\partial r}\right)_L = 0,
 \end{equation}
 or in terms of the gravitational potential $\Phi$,
 \begin{equation}
 \label{potentials-Newton}
 \left( \frac{d \Phi}{d r}\right) -
 L^2\left( \frac{1}{r^3}\right) = 0.
 \end{equation}
 Let us consider almost circular motion of particles on the
 $\theta =\pi/2$ ``equatorial'' plane in the Schwarzschild
 spacetime. From the radial component of the four-velocity
 $u^r$ let us construct
 a positive small quantity $V^2 \equiv (u_r)^2\,g^{rr} \ll 1$.
 It is known that $u_t$ and $u_{\phi}$ are constants of motion,
 therefore $L \equiv - u_{\phi}/u_t$ is also a constant of motion
 (the specific angular momentum).
 The $1 = u_i\,u_k\,g^{ik} = (u_t)^2g^{tt} +
 (u_{\phi})^2g^{\phi\phi} + (u_r)^2g^{rr}$
 condition may be written in the form,
 \begin{equation}
 \label{velocity-unity}
 \frac{1}{2}\ln(1 + V^2) = \ln (u_t) +
 \frac{1}{2}\ln \left[ g^{tt} +
 L^2\,g^{\phi\phi}\right]
 \end{equation}
 Expansion of the left-hand side yields $V^2/2$. One also defines
 $E \equiv \ln(u_t)$ and,
 \begin{equation}
 \label{effective-einstein}
 U(r, L) \equiv -\frac{1}{2} \ln \left[ g^{tt} +
 L^2\,g^{\phi\phi}\right].
 \end{equation}
 This brings equation (\ref{velocity-unity}) into a form identical
 with the Newtonian formula (\ref{newton-effective-equation}).
 Thus, the Newtonian condition (\ref{circular-Newton}) for the
 vanishing derivative of the effective potential may be applied to
 the relativistic effective potential (\ref{effective-einstein}),
 which gives
 \begin{equation}
 \label{einstein-circ-condition}
 \left(\frac{d g^{tt}}{d r}\right) +
 L^2 \left(\frac{d g^{\phi\phi}}{d r}\right) = 0.
 \end{equation}
 Because at the equatorial plane $g^{\phi\phi} = -1/r^2$,
 and $g^{tt} = r/(r - r_{\rm
 G})$, this may be written in the form,
 \begin{equation}
 \label{pacz-wiita-einstein}
 \frac{d}{d r}\left( -\frac{GM}{r - r_{\rm
 G}} \right) - L^2 \left(\frac{1}{r^3}\right) = 0.
 \end{equation}
 Comparing Newton's equation (\ref{potentials-Newton}) with
 Einstein's equation
 (\ref{pacz-wiita-einstein}), we see that the gravitational
 potential in both equations has to have the same
 Paczy{\'n}ski-Wiita form (\ref{pacz-wiita-potential}).
 In deriving equation (\ref{pacz-wiita-einstein}) we used,
 \begin{equation}
 \label{by-a-constant}
 g^{tt} = \frac{r}{r - r_{\rm G}} = \frac{r_{\rm G}}{r - r_{\rm
 G}} + 1 = 2\,\Phi(r) + 1.
 \end{equation}
 Thus, the Keplerian angular
 momentum derived (in the Schwarzschild spacetime) according to
 Einstein's theory, and derived with the Paczy{\'n}ski-Wiita
 potential are both given by the same formula,
 \begin{equation}
 \label{angular-momentum}
 L_{K}^2 = \frac {G M r^3}{(r - r_{\rm G})^2}.
 \end{equation}
 In Newton's theory the angular momentum $L$ and angular velocity
 $\Omega$ are connected by $L = \Omega r^2$, but in Schwarzschild
 geometry by, $L = \Omega r^2/(1 - r_{\rm G}/r)$.
 Therefore, the Keplerian angular velocity calculated in
 Schwarzschild geometry and in Paczy{\'n}ski-Wiita potential are
 {\it not} the same,
 \begin{equation}
 \label{angular-not-equal}
 \Omega_{SCH} = \left(\frac{GM}{r^3}\right)^{1/2} \not=
 \left(\frac{GM}{r^3}\right)^{1/2}\left( \frac{r}{r -r_{\rm G}}\right) =
 \Omega_{PW}.
 \end{equation}

\section{Other potentials, special relativity}

 \cite{wagoner} found that the potential given by a fitting
 formula
 \begin{equation}
 \label{angular-wagoner}
 \Phi_{NW} = -\left(\frac{GM}{r}\right)\left[ 1 - 3\frac{GM}{r\,c^2} +
 12\left(\frac{GM}{r\,c^2}\right)^2\right]
 \end{equation}
 reproduces the angular velocity $\Omega(r)$ and the radial
 epicyclic frequency $\kappa(r)$ {\it better} than the
 Paczy{\'n}ski-Wiita potential. The fitting formula used by
 \cite{kluzniak} reproduces the ratio $\kappa(r)/\Omega(r)$ of
 these frequencies {\it exactly}:
 \begin{equation}
 \label{angular-kluzniak}
 \Phi_{KL} = \left(\frac{GM}{3r_{\rm G}}\right)\left[1
 - {\rm e}^{3r_{\rm G}/r} \right]
 \end{equation}
 \cite{karas} discussed the Newtonian potential suitable for modeling
 the gravity of the Kerr black hole, including the Lense-Thirring
 effect.

 Neither any of these three potentials, nor a few other potential
 introduced by some other authors, became popular. Nowadays much more
 astrophysicists know Einstein's general
 relativity than in the late 1970s, but quotations to Paczy{\'n}ski-Wiita
 potential show no sign of decline\footnote{Number of quotes in years 2000-2008,
 according to ADS: 21, 40, 32, 37, 45, 39, 30, 37, 46.}.

 Velocities of matter calculated with the Paczy{\'n}ski-Wiita potential
 could exceed the light speed. This creates a serious problem
 when one calculates the observed appearance of matter (e.g. spectra)
 by the method of ray tracing. \cite{special-relat} found a
 fix for this problem by showing how to incorporate the effects of
 special relativity into the Paczy{\'n}ski-Wiita scheme: one should
 interpret the ``true'' physical velocities
 in terms of the calculated ones by $V_{cal} = V_{tru}/(1 -
 V_{tru}^2/c^2)$. Here $V_{(...)}$  denotes each of the three
 components of the velocity, i.e. $V_r, V_{\theta}, V_{\phi}$.


 \section{Conclusions}

 The Paczy{\'n}ski-Wiita potential (\ref{pacz-wiita-potential})
 accurately models in the Newtonian
 theory general relativistic effects that determine motion of matter
 near a non-rotating black hole.

 The Paczy{\'n}ski-Wiita potential is neither an approximation
 of relativistic gravity, nor a fitting formula. Instead, it
 is a unique (``pseudo'') Newtonian model of the gravity of a
 non-rotating black hole. It reproduces {\it exactly}:
 \begin{itemize}
 \item The location of the marginally stable orbit $r_{ms} =$
 ISCO.
 \item The location of the marginally bound orbit $r_{mb}$.
 \item The form of the Keplerian angular momentum $L(r)$.
 \end{itemize}
It also reproduces accurately, but not exactly, the form of the
Keplerian angular velocity $\Omega(r)$, and of the radial
epicyclic frequency $\kappa(r)$.


\vfill


\begin{acknowledgements}
      I acknowledge support from
Polish Ministry of Science grant N203 0093/1466 and Swedish
Research Council grant VR Dnr 621-2006-3288
\end{acknowledgements}


\end{document}